# A complex multi-notch astronomical filter to suppress the bright infrared sky

J. Bland-Hawthorn,[1,2] S.C. Ellis,[1,3] S.G. Leon-Saval,[1,2] R. Haynes,[3,4] M.M. Roth,[4] H.-G. Löhmannsröben,[5] A.J. Horton,[3] J.-G. Cuby,[6] T.A. Birks,[7] J.S. Lawrence,[3,8] P. Gillingham,[3] S.D. Ryder,[3] C. Trinh.[1]

**A long-standing and profound problem in astronomy is the difficulty in obtaining deep near-infrared observations due to the extreme brightness and variability of the night sky at these wavelengths. A solution to this problem is crucial if we are to obtain the deepest possible observations of the early Universe since redshifted starlight from distant galaxies appears at these wavelengths. The atmospheric emission between 1000 nm and 1800 nm arises almost entirely from a forest of extremely bright, very narrow hydroxyl emission lines that varies on timescales of minutes. The astronomical community has long envisaged the prospect of selectively removing these lines, while retaining high throughput between the lines. Here we demonstrate such a filter for the first time, presenting results from the first on-sky tests. Its use on current 8m telescopes and future 30m telescopes will open up many new research avenues in the years to come.**

## INTRODUCTION

At visible wavelengths, the terrestrial night sky is very dark which allows astronomers to see back to within a few billion years of the Big Bang using the most powerful optical telescopes on Earth. At infrared wavelengths, the night sky is orders of magnitude brighter due to hydroxyl emission in the upper atmosphere[1]. This is a fundamental obstacle because a great deal of information about the early universe emerges in this part of the spectrum. One approach to bypassing the atmosphere is to launch a space telescope to get above it. This option, however, is hugely expensive and the size of an orbiting telescope is limited.

A solution to the apparently insoluble problem of the infrared night sky emerged in 2004 when our team began to explore new developments in photonics. The infrared sky is bright because the atmosphere glows in hundreds of very narrow spectral lines. Without these lines, the night sky would appear 30-60 times darker. So how do we suppress such a large number of irregularly spaced night-sky lines by factors of hundreds to thousands in an efficient way? Previous attempts using ruled gratings and masks[1,2] are ultimately flawed because of scattering by the grating and bulk optics[3]. The hydroxyl emission must be filtered before the light is allowed to enter the spectrograph.

Two technological innovations were required to achieve an efficient sky-suppressing filter. The first was non-periodic fibre Bragg gratings[4,5] (FBG) capable of suppressing up to 400 narrow lines at high resolution ($\lambda/\delta\lambda \approx 10{,}000$) and high attenuation (~30 dB) over a large bandpass (~200 nm) with low attenuation between the lines (<0.2 dB). The second innovation was a multi-to-single mode fibre converter,[6,7,8] a device we call a "photonic lantern." Multimode fibres are needed to collect light that has been smeared by atmospheric turbulence; single mode fibres are required for the gratings. The converters provide efficient interchange between these two formats.

Here we demonstrate that our new complex grating technology solves the problem of suppressing the bright infrared night sky. The bright lines are completely removed with high efficiency and at high spectroscopic resolution in order to minimize the loss of spectral coverage. This is a prototype technology that we propose to streamline in the near future in order to foster widespread use on future astronomical instruments.

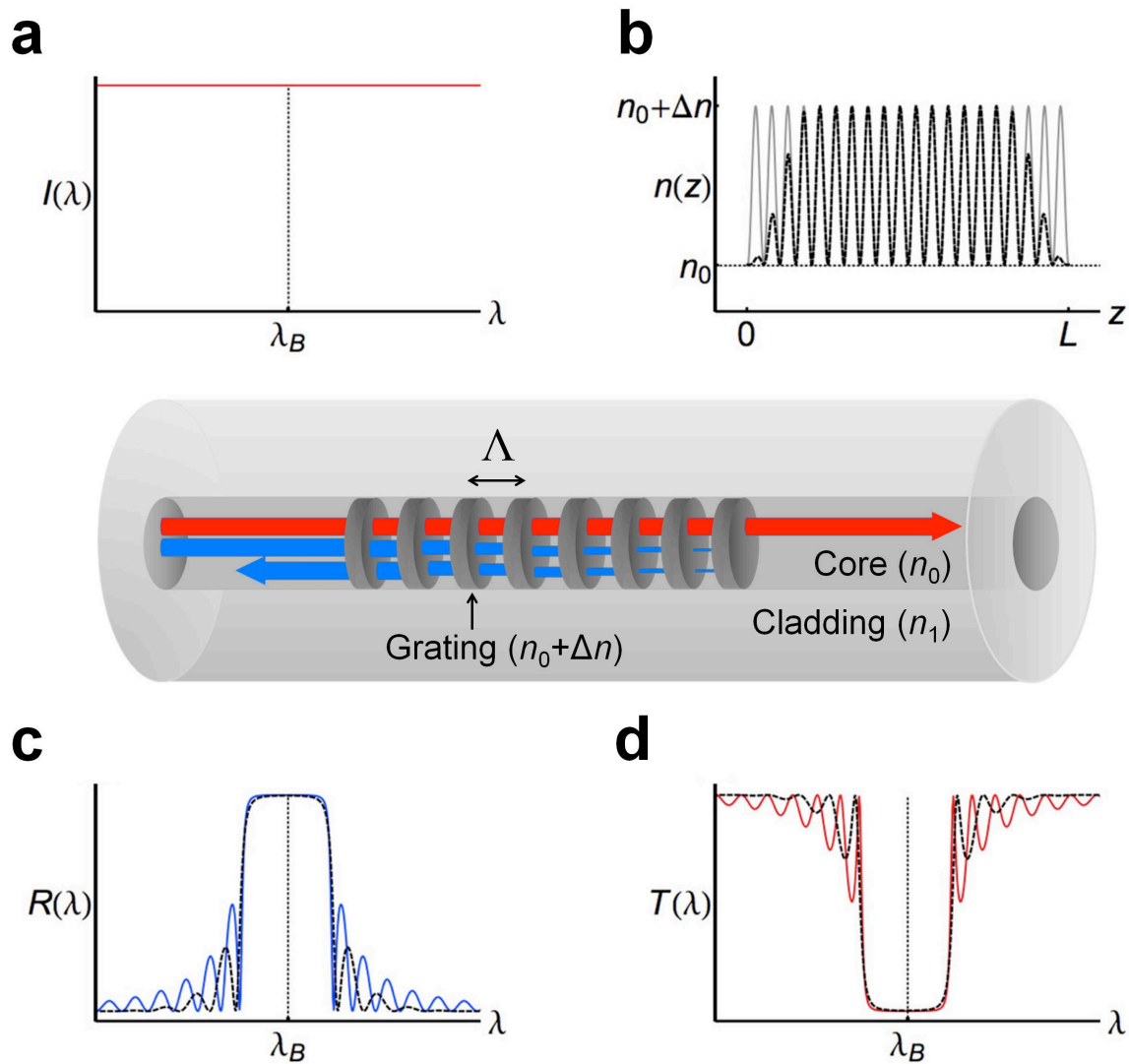

Fig. 1: **The operation of the fibre Bragg grating (FBG)**. The transmitted light (red arrow) enters the grating from the left; the reflected light (blue arrow) is also shown: (a) the input spectrum $I(\lambda)$ centred at the Bragg wavelength $\lambda_B$; (b) the refractive index profile $n(z)$ of the FBG showing both an unapodized (light grey) and an apodized (black) profile; (c) the reflected spectrum $R(\lambda)$ and (d) the transmitted spectrum $T(\lambda)$ showing the unapodized (solid) and apodized (dashed) responses.

## RESULTS

### Fibre Bragg grating

At the present time, the FBG is the only viable technology for achieving an optical filter with many non-periodically spaced notches[4,5] and a high degree of suppression (up to 30 dB in transmission) over a very broad spectral band. Further requirements are that each notch must be rectangular and define a narrow wavelength interval ($\delta\lambda$=0.16 nm such that $\lambda/\delta\lambda \approx$ 10,000), and the inter-notch transmission must be better than 90%.

The principle of the FBG is that light propagating along an optical fibre can be made to undergo Fresnel reflections at many refractive index increments printed onto the fibre core. If the modulation describes a grating, light can be made to reflect back along the full length of the grating. The grating is defined by a complex phase and modulation amplitude along its length. The complex filter observed in transmission arises from interference between the forward-propagating field and the backward-propagating field across the grating. For these reflections to add up coherently, the fibre must be single-moded (i.e. propagation vector aligned with the fibre axis). In a multimode fibre, a discrete wavelength has $M$ spatial modes of propagation where $M = V^2/4$ (not counting a factor of two arising from the polarization of light). The so-called $V$-parameter is given by $V = \pi d$ NA$/\lambda$ where $d$ is the fibre diameter and NA is its numerical aperture. (The formula $M = V^2/\pi^2+1$ occasionally quoted in text books is incorrect.) Each mode experiences the grating along a different wavevector which leads to the notches being smeared out.[4]

### Basic principles

Consider the simplest case of a sine-wave modulation in refractive index along the fibre with grating period $\Lambda$. As we describe under the Methods section, this can be readily achieved by exposing the Ge-doped core to a UV hologram. The reflection efficiency is maximised at the Bragg wavelength $\lambda_B = 2n_o\Lambda$ where $n_o \approx 1.5$ is the refractive index of the fibre core and $\Lambda$ is the grating period (see Fig. 1). For a practical grating, we must specify the grating length $L$ and the modulation amplitude $\Delta n$. If $\Delta n$ is too high, light does not propagate far into the grating; if it is too low, light does not reflect. The strongest response in the narrowest possible notches occurs in the *weak* grating limit for which[9]

$$\left(\frac{\delta\lambda}{\lambda}\right) = \frac{1}{2}\sqrt{\left(\frac{\Delta n}{2n_o}\right)^2 + \left(\frac{\Lambda}{L}\right)^2} \qquad (1)$$

From this equation, we can determine the required modulation amplitude ($\Delta n \sim 10^{-4}$) and minimum grating length for which we have adopted $L = 50$ mm. The peak reflectivity of the grating, $R_g = \tanh^2 \varkappa L$, is determined by the grating amplitude $\varkappa$ (in units of inverse length) and the grating length $L$. The grating amplitude is related to the induced refractive index modulation $\Delta n$ along the fibre axis $z$ by $\varkappa(z) = \pi\, \Delta n(z)/(2\Lambda(n_o + \langle \Delta n \rangle))$ where $z$ is the distance along the FBG. The variable $\varkappa$ is the coupling efficiency of the propagating and counter-propagating electric fields and defines the grating strength. The quantity $\langle \Delta n \rangle$ is the average refractive index change within the grating modulation.

Fig. 1 illustrates the refractive index profile and response in transmission of a fibre Bragg grating defined by the parameters ($L$, $\Lambda$, $n_o$, $\Delta n$). This illustrates several things. The notch is far from rectangular and has finite width; the wings of the notch are too strong and they exhibit ringing; there is a great deal of redundant information in the original grating specified by $L/\Lambda \sim 10^5$ data points. All of these properties of the simple grating design are readily understood in terms of Fourier transform theory.[9] A well known result is that the wings can be greatly suppressed, and the notch can be squared off, by shaping (apodizing) the upper envelope of the refractive index profile (see Fig. 1).

The example above illustrates that a great deal of information can be embedded within an FBG design, orders of magnitude more than is found in a monolithic interference filter with ~100 layers built up from two or more materials. So how do we arrive at a multi-notch design unevenly spaced in wavelength that operates over a broad spectral window with the highest possible transmission? One approach we have used is to print discrete notch designs side by side along the grating, but this quickly breaks down.[4] It is extremely labour intensive to move the fibre along in stages under controlled conditions and to print the next notch, particularly when our basic design calls for $N\sim100$ notches and multiple FBGs.

## Ultra broadband fibre Bragg grating

We must consider extremely complex grating structures in order to make a practical sky-suppressing filter. The entire grating contributes important information to all of the notches which requires a high level of manufacturing stability during the course of an hour-long exposure. Even greater flexibility in the FBG design is achieved by incorporating phase variations, in addition to refractive index modulations, along the grating length. We start with a general description of the grating structure, i.e.

$$n(z) = n_o + \frac{\Delta n(z)}{2}\left(1 + \sin(\frac{2\pi}{\Lambda}z + \Delta\phi(z))\right), \qquad 0 \leq z \leq L \qquad (2)$$

where $z$ is the physical length along the fibre axis, the term $(2\pi/\Lambda)z$ describes the phase delay at each reflecting plane in the FBG (Fig. 1), and the parameter $\Delta\phi$ describes the phase variations (dephasing) with respect to the linear term. Just how $\Delta n$ and $\Delta\phi$ variations are printed onto the FBG is described under the Methods section below. In Fig. 1, the simple grating uses a constant $\Delta n$ while the apodized grating adopts a "raised cosine" envelope in $\Delta n$ to suppress the reflections at the extremes of the grating; for both gratings, $\Delta\phi = 0$.

A full discussion of our aperiodic grating design is highly technical and discussed in detail elsewhere.[5,10] In broad summary, we must solve the coupled mode equations describing light propagation in an FBG[10],

$$\frac{\partial E_b}{\partial z} + i\delta E_b - q(z)E_f = 0$$
$$\frac{\partial E_f}{\partial z} - i\delta E_f - q(z)E_b = 0 \quad (3)$$

where $E_f$ and $E_b$ are the amplitudes of the forward and backward propagating fields. Equation (3) constitutes a pair of non-linear partial differential equations that can be solved using direct/inverse scattering transforms. (This method is due to Lax[12] and was originally developed to solve non-analytic wave equations.) The grating function $q(z)$ is now expressed in its complex form such that $q(z) = \varkappa(z) \exp[i\phi(z)]$, where the amplitude function $\varkappa = |q|$ and the phase function $\phi = \mathrm{Arg}(q)$. The coupled equations must satisfy all wavelengths in a broad spectral band $\Delta\lambda$ centred at $\lambda_o$ set by the width $w_o$ of the UV laser beam, i.e. $w_o = \lambda_o^2/2n_o\Delta\lambda$. The wavelength dependence enters through the variable $\delta$ defined as the wavelength (detuning) offset normalized to $\lambda_o$. The transmission and reflection profiles, $T(\lambda)$ and $R(\lambda)$, are computed through a matrix transformation.[9]

To arrive at an optimal filter design, first note that the discrete set of $N$ notches is defined in the wavelength domain. Each of these requires a refractive index modulation in the physical domain that is as distinct as possible from the others. Each notch can be thought of as a "subgrating" described by a non-linear discrete Fourier transform $s_m(k) = |s_m(k)| \exp(i\Phi(k))$ defined in terms of the wavenumber $k = 2\pi n_o/\lambda$ ($m = 1, N$). If we were to simply add the $s_m$ transforms together, there would be strong interference in transmission between the notches leading to an unusable FBG filter. The notch positions are fixed in wavelength and therefore in $k$, but the phase offset $\Phi$ is arbitrary and lies at the core of our iterative scheme. Moreover, we require that every notch in $T(\lambda)$ be as squared off as possible. (The apodisation suppresses the wing behaviour but does little for the core structure of the notch.) The target function for the filter shape has the following form:

$$|s_m(k)| = \sqrt{\frac{R_m}{\cosh[2(k-k_m)/\Delta k_m]^4}} \quad (4)$$

where $R_m$ is the peak reflectivity of the $m$-th channel, $k_m$ is its centre and $\Delta k_m$ is its width. By good fortune, this highly desirable reflection profile greatly aids convergence for complex multi-channel filter designs.

### Grating design

For the optimisation procedure, a key insight comes from noting that the phase $\Phi$ of a notch can be expanded about its centre through a Taylor series, such that

$$\Phi_m(k) = \Phi_m^0 + d_1 k + \frac{1}{2!} d_2 k^2 + O(k^3) \tag{5}$$

Each term has a clear physical meaning. The constant $\Phi^0$ is the relative phase (dephasing angle) of each notch with respect to the other notches. The linear term $d_1$ is the physical shift of the notch pattern along the fibre to dephase it as much as possible from all other notches; the quantity $d_2$ is the dispersion coefficient which is a constant of the fibre material and contributes only weakly to the design. A final grating design is achieved through optimising the set of dephasing angles and displacements that minimize the interference between the notches. The Taylor expansion has reduced the optimisation problem from $10^5$ to $2N$ data points, or $3N$ data points if the notches are to have different depths.

There are experimental constraints. The optimisation must minimise the maximum amplitude of the reflection grating to a practical threshold of $\varkappa_{max} \approx$ 2-3 mm$^{-1}$ (which ensures the maximum information content and the highest coupling efficiency) while keeping the grating length to 50 mm set by the UV holographic interferometer. As a result of reducing the interference between the notches in the iterative procedure, the complex envelope[5] of the grating profile $q(z)$ from the sum of $N$ sub-gratings becomes smoother and stays below the maximum coupling efficiency $\varkappa_{max}$. The rms departure from a smooth envelope is our primary figure of merit for converging on a practical design. The multi-channel grating profile $q(z)$ is generated from

$$q(z) \approx \sum_{m=1}^{N} G^{-1}[s_m(k)] \tag{6}$$

where $G^{-1}[.]$ is the inverse transform scattering operator. The functions $s(k)$ and $q(z)$ are highly non-linear with respect to each other. Thus the coupled equations in (3) present an inverse scattering problem which must be solved iteratively in order to arrive at a practical grating design $q(z)$. The operational details of how this complex optimisation is carried out subject to the above constraints is discussed elsewhere[5,10]; these references also provide examples of the complex grating amplitudes and phase functions that result from the optimisation.

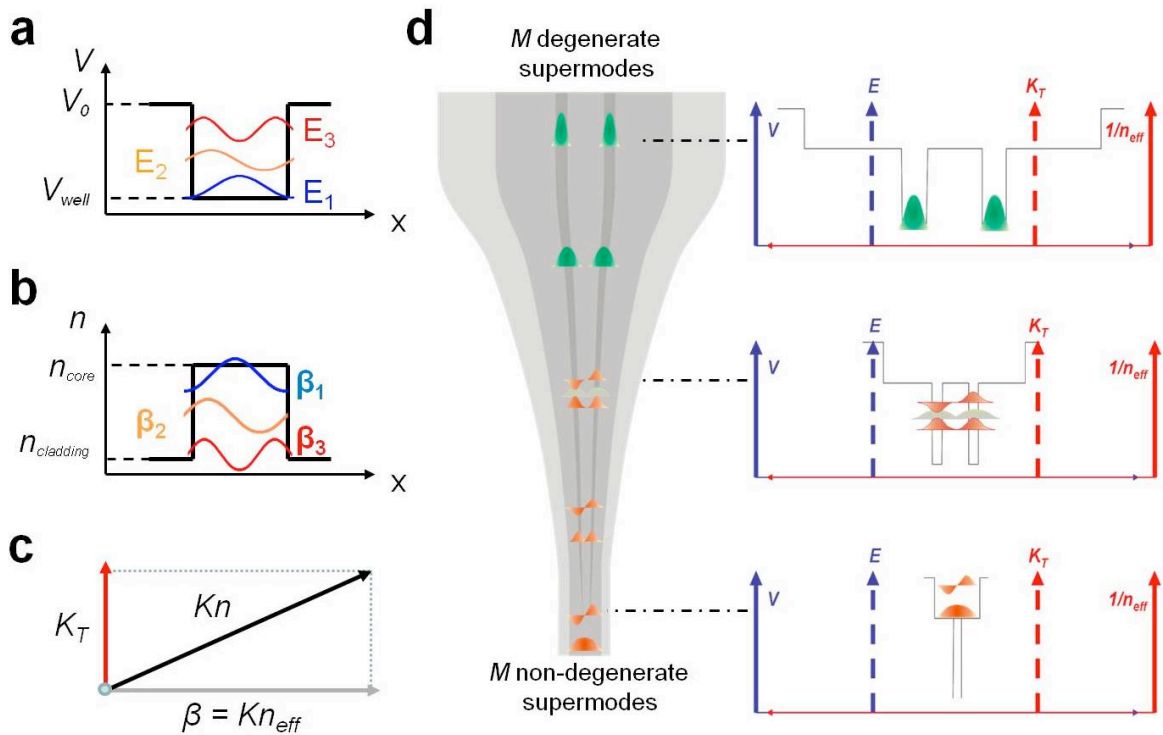

Fig. 2: **How the photonic lantern works.** Schematics of (a) a one dimensional quantum well, and (b) a one dimensional waveguide illustrating the close analogy between the propagating modes in an optical fibre ($\beta_1$, $\beta_2$, $\beta_3$) and the energy levels of a quantum well ($E_1$, $E_2$, $E_3$). (c) The $Kn$ wavevector is made up of a transverse component $K_T$ and a mode propagation constant $\beta$ defined in terms of the effective refractive index seen by the mode. (d) Here we illustrate the function of a photonic lantern taper starting at the input bundle of $M$ single mode fibres (top). Initially the strongly guided radiation in each core is uncoupled; upon entering the adiabatic taper, the radiation is increasingly guided by the cladding. In a perfect system, this leads to $M$ supermodes that evolve to become $M$ electromagnetic modes at the multimode fibre output (bottom). This has a close analogy to what happens when we bring atoms close together in the Kronig-Penney model of quantum mechanics. Initially, the electrons are confined but as the atoms are squeezed together, the electrons tunnel out and propagate freely in the bulk.

## Photonic lantern

The photonic lantern, first demonstrated[6,7,8] in 2005, features an array of single-mode fibres (SMF) surrounded by a low index layer that is adiabatically tapered down to form a multimode fibre (MMF) on input or output depending on the intended direction (Fig. 2). Efficient coupling is achieved in both directions if the number of (unpolarized) excited modes in the MMF is equal to the number of SMFs in the bundle. Light can couple between the bundle of SMFs and the MMF via a gradual taper transition. If the transition is lossless, then the supermodes (group of the degenerate independent SMF modes) of the SMF bundle evolve into the modes (group of non-degenerate supermodes) of the MMF core, and vice versa (Fig. 2). The second law of thermodynamics does not allow lossless coupling of light from an arbitrarily excited MMF into *one* SMF, but if the MMF has the same number of degrees of freedom as the SMF bundle, then lossless (adiabatic) coupling becomes possible by conserving the entropy of the system.

Just how these *M* uncoupled SMF electromagnetic modes evolve through an adiabatic taper to become the *M* electromagnetic modes of the output MMF can be appreciated by analogy with the Kronig-Penney model[9,13] for the interaction of electrons in a periodic potential well. We can compare the photonic transition between an isolated bundle of *M* SMFs and a MMF with *M* modes to a Quantum Mechanic (QM) system which evolves from *M* isolated potentials, each with a single discrete allowed energy level, to a single isolated broader potential with *M* discrete energy levels. In order to make this analogy, we sketch the 1D refractive index profile of a step-index fibre, but using $1/n$ instead of $n$ in the vertical axis (Fig. 2). This simple inversion allows the refractive index profile of our photonic lantern to be represented in the same fashion as the 1D Kronig-Penney model of periodic potentials. We can therefore compare optical fibre cores with quantum wells as shown in Fig. 2(a) and (b).

In the QM case, energy is used to define the wavefunctions corresponding to discrete energy levels; however this is not applicable to the electromagnetic (EM) case of spatial modes propagating along a fibre core. The spatial modes in the EM case have the same energy $E$ and are distinguished by their propagation constant $\beta = Kn_{eff}$ ($K$ being the wavenumber and $n_{eff}$ the effective index of the mode) and a transverse wavevector $K_T$ (see Fig. 2(c)). In our analogy $K_T$ (EM) and $E$ (QM) behave qualitatively the same. We compare the change in energy of the standing wave solutions of the electron inside the quantum well with the change in $K_T$ of the spatial modes in the waveguide. A fibre core can be designed to have only one spatial mode by tailoring the refractive index profile. This mode has the highest $\beta$ and its electric field concentrated in the region of highest $n$, hence the highest mode effective index ($n_{eff}$) value and the lowest transverse wavevector ($K_T$) (see Fig. 2(c)). In the QM case of an isolated potential well, only discrete energies for the electron wavefunction are allowed, and the electron wavefunction takes the form of standing waves. With the right potential and geometry, a potential well can be designed to allow only one discrete energy level (i.e. the ground state). These standing wave solutions of the independent quantum wells have the lowest energy ($E$) and typically their amplitude is concentrated in the regions of lowest potential ($V$), and these can be considered the fundamental modes.

At the start of the transition (Fig. 2(d)), each quantum well allows only one electron in its lowest energy state (fundamental mode). The taper transition renders the quantum wells progressively shallower such that each electron begins to tunnel through its barrier. With the wells closer together, the leaky "conduction" electrons behave as if confined to a periodic crystal. At the point where the taper ends, the wells have essentially vanished, and the collective behaviour of the electrons is described by *M* standing waves (cf. supermodes) confined to a single broad potential well (cf. multimode core).

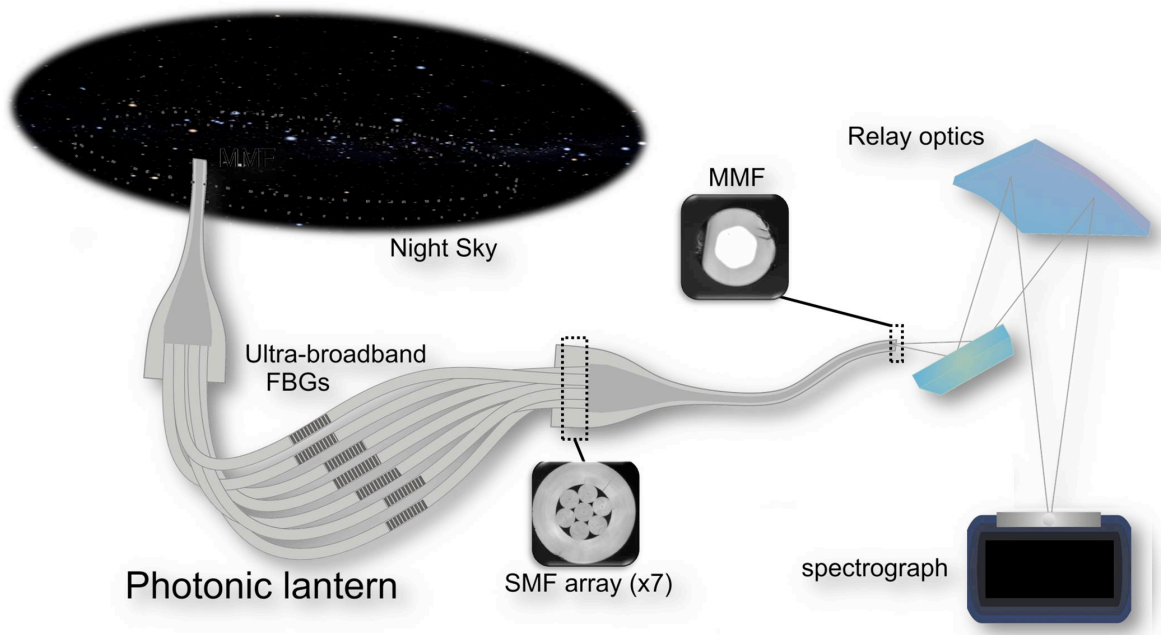

Fig. 3: **The set up for the on-sky tests of the sky-suppressing fibre.** Light from a multimode fibre exposed to the night sky is fed through a photonic lantern and a parallel array of sky-suppressing FBG filters. The comparison fibre (not shown) has an identical layout except the photonic lantern does not include FBGs. The seven reflection gratings placed between the two adiabatic tapers have identical performance characteristics. The output multimode fibre (MMF) illuminates relay optics, which redirect the light into an infrared spectrograph.

Experiment

The sky-suppressing fibres were tested on the nights of 17-19 December 2008 at Siding Spring Observatory in New South Wales, Australia. The choice of location gave us access to an infrared spectrograph; no telescope was used. The end faces of two 60μm core fibres were illuminated with light from a ~10 degree patch of sky. Both fibres underwent a taper transition to 7 hexagonally packed single-mode fibres before undergoing another taper transition in reverse back to a multimode fibre (Fig. 3). In one lantern, identical gratings tuned to the 63 brightest OH lines within 1440-1630 nm were printed into the 7 single mode fibres. The other lantern acted as a control, with no gratings. The light from these fibres was relayed into the IRIS-2 infrared spectrograph. Exposures of 900s were taken simultaneously through both fibres, consisting of 31 sequential reads of 30s each in order to minimize detector read noise and to reject charged particle events. The IRIS-2 spectral resolution imposed by the sapphire grism was $\lambda/\delta\lambda \approx 2400$.

DISCUSSION

In Fig. 4, the results demonstrate the extraordinary power of photonic OH suppression where, for example, no residuals are seen at the locations of the brightest lines. Strong residuals are a feature of essentially all infrared spectroscopic studies in astronomy. The OH lines are suppressed at a resolution four times *higher* than is seen in Fig. 4(c). Between the lines, the throughput is high as shown by the OH lines redward of the grating limit. There is a minor insertion loss in our prototype due to imperfect mode matching within the photonic lantern (see Methods). Devices now in development will be capable of converting 60-120 spatial modes to the same number of single-mode tracks allowing for OH suppression within ~100μm core fibres. Our latest grating designs[5] suppress up to 400 OH lines across the 1000-1800 nm window resulting in a background 30-60 times fainter than is possible today across the near infrared window.

The success of our on-sky demonstration shows that the problem of the bright near-infrared night sky has finally been solved. *The prototype constitutes the most complex optical filter ever constructed*. Its application on current 8m telescopes and future 30m telescopes will open up many new research avenues in the coming years. In astronomical instruments, we propose to form an image by constructing a mosaic of MMFs fed by a microlens array[13]. But to do this, we must use a different approach. Consider a 30×30 microlens array feeding a matched array of photonic lanterns at f/5, each with a core size of 100 μm. At 1500 nm, the number of spatial modes in each lantern requires 90 FBGs such that the entire array would need 81,000 FBGs independently manufactured and spliced into the lanterns. This is not tenable. The problem is somewhat reduced if the focal plane has been corrected with adaptive optics. A lantern core size of 25 μm reduces the required number of FBGs to 7 but the full array would still need 6,300 FBGs. Our solution involves printing FBGs into multicore fibres with both ends drawn down in the manner of a photonic lantern.[6] These devices are presently an active area of investigation. We believe this will allow mass production of sky-suppressing fibres for more widespread application and more ambitious astronomical instruments.

Looking further into the future, we envisage spectrographs that use photonic components exclusively without the need for bulk optics. The light propagates through a waveguide where it is gathered, dispersed and imaged onto a detector[14-17]. This approach, while challenging to implement in practice, will reduce the cost, size and complexity of existing spectrographs. With such a revolutionary approach, it should be possible to remove thousands rather than hundreds of night sky lines with the aid of a new generation of photonic filters operating at high efficiency and at high spectroscopic resolution.

METHODS
Fibre Bragg Grating

The refractive index modulation pattern in the FBG comes from exposing the photosensitive fibre core to a spatially varying pattern of UV photons produced by the Mach-Zehnder interferometer. Below 300nm, the UV photons break down the SiO bonds thus causing microscopic variations in the refractive index of the medium. The grating in Fig. 3 is printed onto a GeSiO$_2$ fibre with a period of $\Lambda = 0.5\,\mu$m. The fibre core size of 8.5 $\mu$m allows only single mode propagation. The fibre has a 125 $\mu$m cladding diameter and a 250 $\mu$m acrylate buffer diameter. The required filter is achieved with a low $\varkappa$ material ($\approx 10$ cm$^{-1}$) over a long baseline; for our grating, $L \approx 100$ mm and $\langle \Delta n \rangle \approx 10^{-4}$. The wavelength interval for suppression $\Delta\lambda$ is related to the beam psf of the UV laser writing system. The beam psf is given by $z_{psf} = \lambda_o^2/(2n_{eff}\Delta\lambda)$ at an operating wavelength of $\lambda_o$ defining the centre of the band. For our first on-sky demonstration, we adopt $\lambda_o = 1600$ nm and $\Delta\lambda = 200$ nm. The derived physical spot size $z_{psf} = 4$ $\mu$m is a factor of two smaller than what the system could deliver, requiring us to print the spectral band in two parts, each of length $L$.

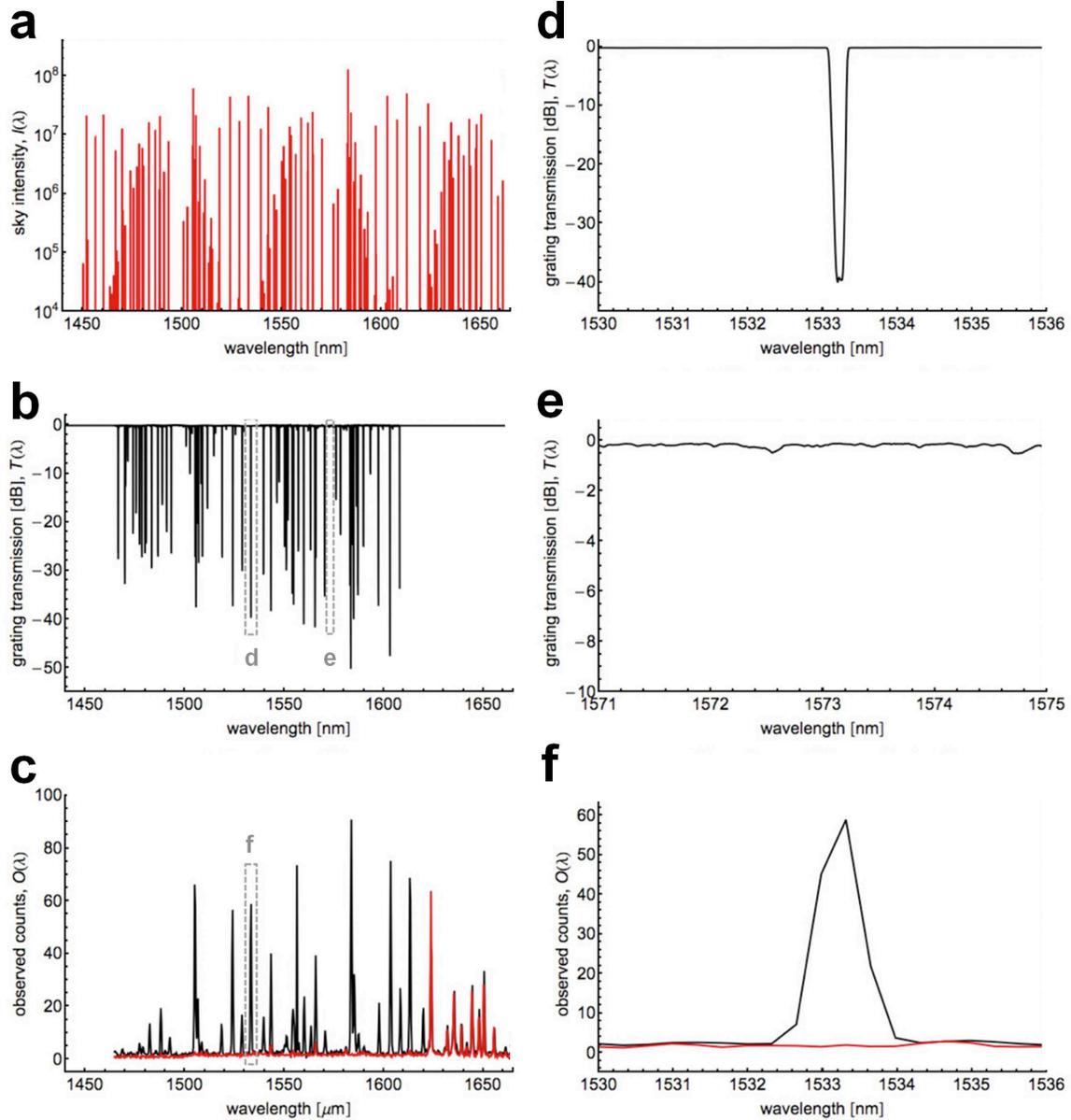

Fig. 4: **Sky spectrum and filter performance.** (a) Infrared night sky model in units of log (phot m$^{-2}$ s$^{-1}$ arcsec$^{-2}$ $\mu$m$^{-1}$). (b) Grating transmission in units of dB measured with an optical spectrum analyser. (c) Comparison of the unsuppressed control spectrum (black) and the suppressed sky spectrum (red). The sky lines are strongly

suppressed in the region covered by the grating. The residual weak continuum is due to starlight in the 10° fibre input beam. The grating region below 1465 nm is blocked by the infrared spectrograph. Zoomed details: (d) The intrinsic width of each rectangular notch is 0.15 nm FWHM; there are no wings and no detectable ringing outside of the notch. (e) The interline losses are <0.15 dB; the vertical scale is now 10 dB to emphasize interline structure. (f) The suppression of each OH sky line is perfect with no detectable residual at the resolution of the spectrograph.

Photonic lantern

The lantern is made by inserting 7 single-mode fibres into a glass capillary tube (Fig. 5), fusing the bundle together into a solid glass element and tapering the element down to a multimode fibre with a core diameter of ~60 $\mu$m and an outer diameter of ~110 $\mu$m. The glass tube has a fluor-doped, low-index layer on the inside of the capillary wall that acts as the cladding of the down-tapered MMF. The core of the MMF consists of the SMFs fused together. The SMFs are OFS Clearlite fibres, with an outer diameter of 80 $\mu$m, a mode field diameter of 7.5 $\mu$m and a single mode cut-off wavelength around 1300 nm. The tube surrounding the fibres is tapered down by a factor of ~3.5. This results in a V-parameter of the original SMFs in the multimode end of ~0.6 at a wavelength of 1550 nm, which means that the light will leak out of the SMFs and be guided by the multimode structure. The tapering of the fibre bundle is done over a length of 40 mm. The tapering of the capillary tube that contains the fibres is done on a GPX-3100 glass processing station from Vytran. By tailoring the amount of heat given to the device during tapering, the point where the fiber bundle is fully collapsed can be controlled. This point is chosen to be where the V-parameter of each SMF is reduced by a factor of ~4. At a distance of 2-2.5 cm into the taper, the cores are so small that they no longer act as individual waveguides. The diameter of the MMF core is ~60 $\mu$m and the numerical aperture of the ring on the inside of the surrounding tube is NA = 0.06. Therefore, the number of spatial modes that can propagate in the MMF is ~13 at wavelengths around 1550 nm. The fabrication method is similar to the "stack-and-draw" technique described in [8].

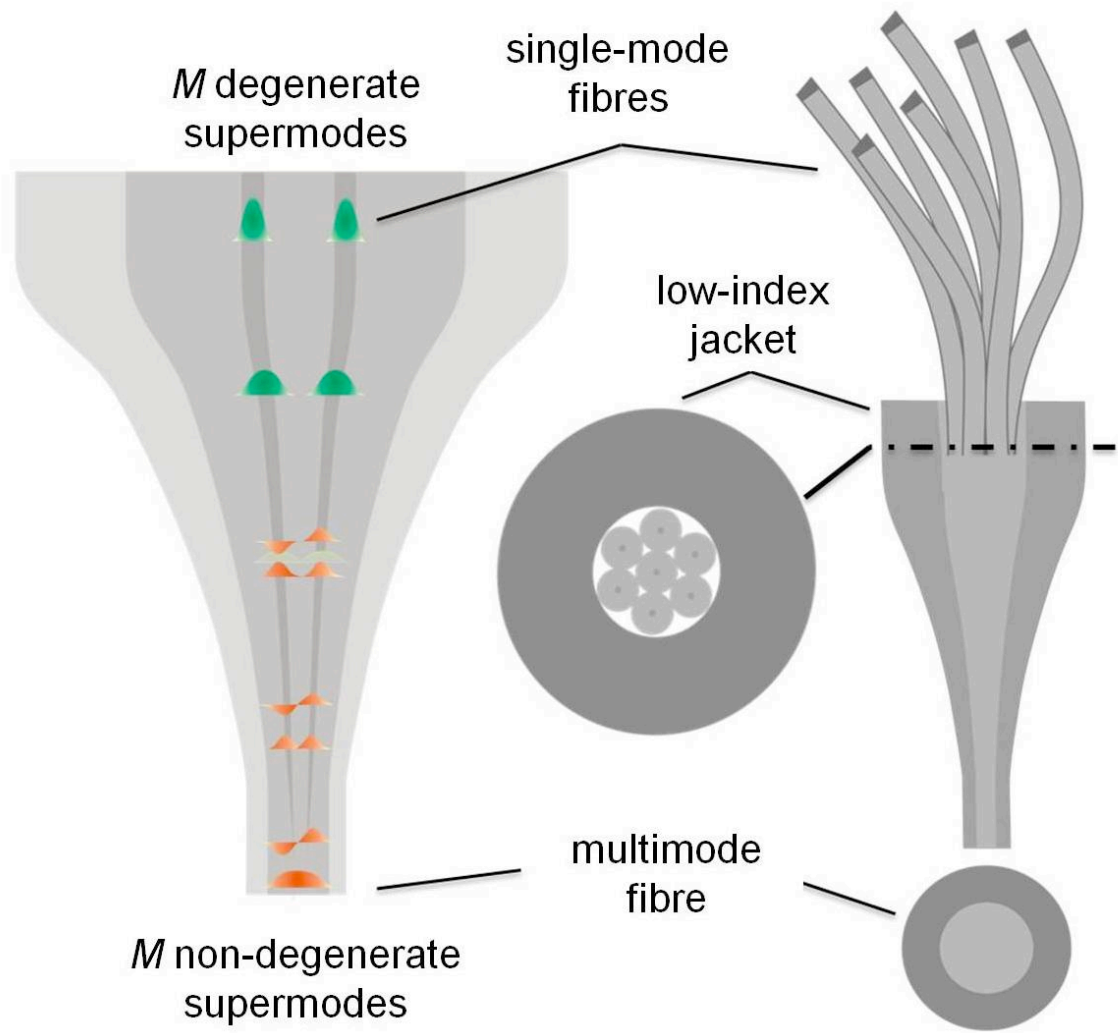

Fig. 5: **How the photonic lanterns are constructed.** A bundle of 7 SMFs are placed inside a low index glass jacket. The choice of the jacket material is very important because it becomes the cladding material for the output MMF after being tapered down. The core material for the output MMF is made from the SMFs (core+cladding) after being tapered down.

Acknowledgments

This research has been supported by the University of Sydney, the Australian Research Council, the Particle Physics and Astronomy Research Council (UK), the Science and Technology Facilities Committee (UK), the Australian Research Council, the Anglo-Australian Observatory, the Konjunkturpaket II (Germany), and the Laboratoire d'Astrophysique Marseille.  JBH acknowledges a Federation Fellowship and SLS acknowledges an Advanced Postdoctoral Fellowship from the Australian Research Council. CT gratefully acknowledges support by the National Science Foundation Graduate Research Fellowship under Grant No. DGE-1035963. We are extremely grateful to the site staff at the Australian Astronomical Telescope for their assistance in setting up the experiment.

Correspondence and request for materials should be addressed to JBH (jbh@physics.usyd.edu.au).

## Author contributions

JBH has overseen all aspects of the OH suppression project. JBH and SLS wrote the manuscript and performed the experiments with SCE, RH, AJH, JSL, PG, SDR and CT. TAB and SLS were involved in the initial photonic lantern development. JGC, MMR and HGL were involved in the grating development.

## Competing financial interests

The authors declare that they have no competing financial interests.

______________________

[1]Sydney Institute for Astronomy, School of Physics, University of Sydney, Camperdown, NSW 2006, Australia

[2]Institute of Photonics & Optical Science, School of Physics, University of Sydney, Camperdown, NSW 2006, Australia

[3]Australian Astronomical Observatory, PO Box 296, Epping, NSW 2121, Australia

[4]innoFSPEC, Leibniz-Institut für Astrophysik, An der Sternwarte 16, D-14482 Potsdam, Germany

[5]Institut für Chemie/Physikalische Chemie, Universität Potsdam, Karl-Liebknecht-Str. 24-25, D-14476 Golm, Germany

[6]Laboratoire d'Astrophysique de Marseille, OAMP, Université Aix-Marseille, France

[7]Department of Physics, University of Bath, Claverton Down, Bath BA2 7AY, UK

[8]Astronomy, Astrophysics & Astrophotonics Research Centre, Macquarie University, North Ryde, NSW 2109, Australia